\documentclass[apj, numberedappendix]{emulateapj}

\usepackage{hyperref}

\newcommand{\dw}{\dot{\omega}}
\newcommand{\teff}{T_{\rm{eff}}}

\begin{document}

\title{A Photometric, Spectroscopic, and Apsidal Motion Analysis \\ 
	of the F-type Eclipsing Binary BW~Aquarii from K2 Campaign 3}
	
\author{Kathryn V. Lester \& Douglas R. Gies}

\affil{Center for High Angular Resolution Astronomy and Department of Physics \& Astronomy, 
\\ Georgia State University, P.O. Box 5060, Atlanta, GA 30302-5060, USA}

\email{lester@astro.gsu.edu, gies@chara.gsu.edu}

\journalinfo{Accepted for publication in The Astronomical Journal} 
\submitted{Accepted 2018 May 4}

% --------------------------------------------------------------------------------------------------------------------------------------
\begin{abstract}
Eclipsing binaries are important tools for studying stellar evolution and stellar interiors. Their accurate fundamental parameters are used to test evolutionary models, and systems showing apsidal motion can also be used to test the model's internal structure predictions. For this purpose, we present a photometric and spectroscopic analysis of the eclipsing binary BW Aquarii, an evolved F-type binary with slow apsidal motion. We model the \textit{K2} C3 light curve using the Eclipsing Light Curve code to determine several orbital and stellar parameters, as well as measure the eclipse times to determine updated apsidal motion parameters for the system. Furthermore, we obtain high-resolution spectra of BW~Aqr using the CHIRON echelle spectrograph on the CTIO 1.5m for radial velocity analysis. We then reconstruct the spectra of each component using Doppler tomography in order to determine the atmospheric parameters.  We find that both components of BW~Aqr are late F-type stars with $M_1 = 1.365 \pm 0.008 \ M_\sun$, $M_2 = 1.483 \pm 0.009 \ M_\sun$,  and $R_1 = 1.782 \pm 0.021 \ R_\sun$, $R_2 = 2.053 \pm 0.020 \ R_\sun$. We then compare these results to the predictions of several stellar evolution models, finding that the models  cannot reproduce the observed properties of both components at the same age. 

\end{abstract}


% --------------------------------------------------------------------------------------------------------------------------------------
\section{Introduction}
The fundamental parameters of eclipsing binary stars are used to test models of stellar evolution by seeking model solutions that fit the observed masses, radii, and effective temperatures at a single age. If a binary orbit is eccentric, it may show apsidal motion, where the periastron position precesses over time due to tidal forces. These tidal forces depend on the internal mass distributions of the stars in the binary system, and therefore provide an additional test of model stellar interiors. This ``apsidal motion test" compares the observed rate of apsidal motion to the predictions of stellar structure models, but requires binary systems with very accurate absolute dimensions (errors $<2\%$) because the tidal forces are also a strong function of the relative radii of the stars \citep{cg93, claret02, cg10}. 

One eclipsing binary system showing apsidal motion is BW~Aquarii ($\alpha_{J2000} = 22:23:15.9,\  \delta_{J2000} = -15:19:56.2$, $V = 10.3$), a pair of late-F stars with an orbital period of $P=6.7$ days and moderate eccentricity. BW~Aqr has a long observational history since it was discovered by Henrietta Leavitt \citep{pickering}. Visual and photographic observations date back to the early 1900s \citep[e.g.][]{robinson68}, while several photoelectric and CCD observations have been obtained more recently for light curve and apsidal motion analyses  \citep{kk86, gronbech87, bulut09}.  The first radial velocity analysis of BW~Aqr was completed by \citet{imbert87}, who combined their spectroscopic results with the photometric results of \citet{kk86} to obtain mass and radius estimates for each component.  They also found that the hotter star is less massive, while cooler star is more massive and evolved to near the terminal age main sequence. BW~Aqr is thus located in an interesting part of the HR diagram where few other systems are found \citep{clausen10},  providing a unique challenge for evolutionary models.

Fortunately, BW~Aqr was observed by \textit{Kepler} during \textit{K2} Campaign 3, providing greater precision and phase coverage than previous photometric observations. We combined this \textit{K2} photometry with newly obtained high-resolution spectra in order to update and better constrain the orbital, physical, and apsidal motion parameters of BW~Aqr. Section~\ref{observations} describes our photometric and spectroscopic observations. Section~\ref{photometry} details the light curve and apsidal motion analyses, while Section~\ref{spectroscopy} explains the spectroscopic analysis. Our results are presented in Section~\ref{results}, where we compare the observed parameters to several stellar evolution models and perform the apsidal motion test. We put our results into context with other studies in Section~\ref{discussion}.  Note that throughout this paper, we refer to the ``primary" as the hotter, less massive star and the ``secondary" as the cooler, more massive star, following the notation of \citet{clausen91}.

% --------------------------------------------------------------------------------------------------------------------------------------
\section{Observations}\label{observations}

\subsection{\textit{K2} Photometry}
BW~Aqr (EPIC 205982900) was observed by \textit{Kepler} from 2014 Nov 15 - 2015 Jan 23 during \textit{K2} Campaign~3 in both short cadence (1 minute) and long cadence (29.4 minute) exposures \citep{howell14}. We downloaded the extracted, long cadence light curve and the short cadence target pixel file from MAST to use in our analysis. The long cadence light curve was produced by the standard \textit{Kepler} pipeline, so we used the \textsc{PyKE} code \citep{pyke} to normalize the Simple Aperture Photometry flux and remove exposures taken during thruster firing as noted in the quality flag. For the short cadence data, we used \textsc{PyKE} to extract the light curve from the target pixel file, subtract the background flux, normalize the light curve, and identify and remove exposures taken during thruster firing. Finally, we converted both the long cadence and short cadence light curves from normalized flux to \textit{Kepler} magnitudes using $K_p=10.233$ \citep{huber16} for the out-of-eclipse magnitude. 

\subsection{Spectroscopy}
We obtained 14 nights of data for BW~Aqr using the CHIRON echelle spectrograph \citep{chiron} on the CTIO 1.5 m telescope during 2015 May 19 - June 19. Three 800 second exposures were taken in fiber mode and averaged for each night, and thorium-argon lamp spectra were taken for wavelength calibration. The CTIO pipeline from Yale University \citep{chiron} was used for all data reduction. We then transformed each spectrum onto a heliocentric, logarithmic wavelength grid and used the reduced flat field spectra for continuum normalization.  The CHIRON spectra cover $4500-8900$\AA \ over 61 orders at an average resolving power of $R \sim 27000$. The average signal-to-noise ratio of our spectra is $S/N = 75$ near the blaze peak of the H$\alpha$ echelle order.

% --------------------------------------------------------------------------------------------------------------------------------------
\section{Light Curve Analysis}\label{photometry}

\subsection{Orbital Ephemeris}
We used the short cadence \textit{K2} C3 light curve for eclipse timing because of its precise, one minute time step. To measure the times of mid-eclipse, we first used a spline interpolation to determine the times of ingress and egress at twenty evenly-spaced depths within each eclipse. We then calculated the time of mid-eclipse from the average times of ingress and egress at all depths. The errors in eclipse time correspond to the standard deviation in the center times from each depth and are on the order of 6-10 seconds, which is not unprecedented. This level of precision has been reached for many other short and long cadence \textit{Kepler} targets with deep eclipses and high signal-to-noise photometry \citep[e.g.,][]{gies12, conroy14, orosz15}. The times of mid-eclipse for the \textit{K2} C3 data are listed in Table~\ref{eclipsetimes}, along with the errors and the eclipse type.  The deeper, primary eclipses are type I and the shallower, secondary eclipses are type II. 

We then combined the \textit{K2} eclipse times with the other eclipse times from the literature \citep{clausen91, bulut09, volkov14}.  Weighted, least squares fits to the primary and secondary eclipse times result in the following linear ephemerides, 
$${\rm Min\ I} \ = {\rm BJD}\ 2456990.46690(2) + 6.7196988(2) E$$  
$${\rm Min\ II}  = {\rm BJD}\ 2456993.65447(2) + 6.7196831(2) E$$
where $E$ is the integer epoch number. The difference in period for the primary and secondary eclipses, albeit small, is indicative of apsidal motion. The true orbital period, known as the sidereal period ($P_s$), corresponds to the average of the periods from the primary and secondary eclipses. For BW~Aqr, we found $P_s = 6.7196909 \pm 0.0000002$ days.

\begin{deluxetable}{cccc}
\tablewidth{0pt}
\tablecaption{Times of Eclipse Minima during \textit{K2} C3 \label{eclipsetimes}}
\tablehead{\colhead{Date} & \colhead{$\sigma$} & \colhead{Eclipse} \\
\colhead{(BJD-2400000)} & \colhead{(days)} & \colhead{Type}  } 
\startdata
 56980.21481     &    0.00044     & II    \\ 
 56983.74726     &    0.00007     & I     \\ 
 56986.93492     &    0.00015     & II    \\ 
 56990.46691     &    0.00007     & I     \\ 
 56993.65449     &    0.00007     & II   \\  
 56997.18658     &    0.00007     & I    \\  
 57000.37420     &    0.00017     & II   \\  
 57003.90621     &    0.00007     & I    \\  
 57007.09379     &    0.00013     & II   \\  
 57010.62606     &    0.00010     & I    \\  
 57013.81352     &    0.00008     & II   \\  
 57017.34571     &    0.00007     & I    \\  
 57020.53319     &    0.00007     & II   \\  
 57024.06555     &    0.00013     & I    \\  
 57027.25252     &    0.00052     & II   \\  
 57030.78506     &    0.00011     & I     \\ 
 57033.97254     &    0.00012     & II   \\  
 57037.50486     &    0.00008     & I    \\  
 57040.69235     &    0.00007     & II   \\  
 57044.22431     &    0.00007     & I    
\enddata 
\end{deluxetable}

\subsection{Light Curve Modeling}
We used the Eclipsing Light Curve (\textsc{elc}) code by \citet{ELC} to model the long cadence light curve of BW~Aqr.  We ran \textsc{elc}'s genetic optimizer to fit for several parameters:  eccentricity ($e$), longitude of periastron ($\omega$) of the primary star, epoch of periastron ($T$),  orbital inclination ($i$), relative radius ($R/a$) of each component, and the effective temperature ratio ($T_{\rm{eff\ 2}} / T_{\rm{eff\ 1}}$). We held the orbital period fixed to the sidereal period, set the orbital semi-amplitudes from spectroscopy (Section~\ref{specorbpar}), and took all limb darkening coefficients  from \citet{vanhamme93}. The model light curve also accounts for the time averaging over the 29.4 minute cadence of the \textit{K2} measurements.

Because BW~Aqr does not show any variations outside of eclipse, such as ellipsoidal variations or reflection effects, the out-of-eclipse points do not hold any information about the orbital parameters of the system. We therefore used only the observed data points during primary or secondary eclipses, then kept only every third data point in order to reduce computation time. This resulted in 174 points used for fitting and 167 degrees of freedom ($\nu$), so we rescaled the optimizer's output $\chi^2$ values such that $\chi^2_{min} = \nu$.  The $1\sigma$ errors in each parameter were calculated by fitting a parabola to the projected $\chi^2$ values and taking $\chi^2 \le \chi^2_{min} + 1$.

\begin{figure}[h!]
\centering
\epsscale{1.2}
\plotone{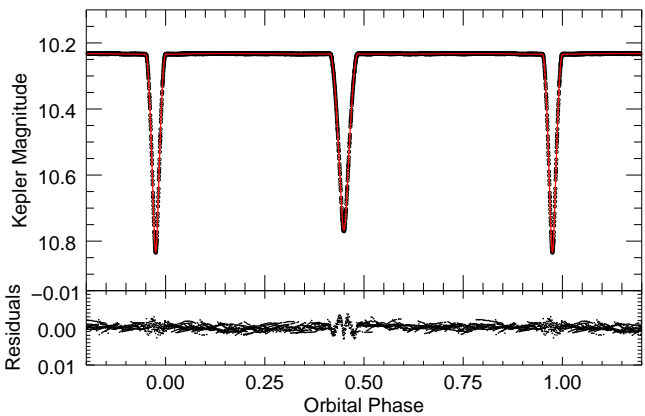} 
\caption{Observed \textit{K2} C3 long cadence light curve (open circles) and best fit Eclipsing Light Curve model (red line). The orbital phase is relative to the epoch of periastron. Residuals of the model fit are shown in the bottom panel. \label{LC}}   
\end{figure}

The best fit orbital parameters are listed in the beginning of Table~\ref{orbpar}, along with the spectroscopic orbital elements found in Section~\ref{specorbpar}. The best fit relative radii are $R_1/a = 0.0840 \pm 0.0010$ and $R_2/a = 0.0966 \pm 0.0009$, and the best fit temperature ratio is $T_{\rm{eff\ 2}} / T_{\rm{eff\ 1}}=0.9930 \pm 0.0016$. The surface gravity of each star was estimated in \textsc{elc}  to be $\log g_1= 4.069$ and $\log g_2 = 3.986$, which are consistent with the findings of \citet{clausen91}.  Figure~\ref{LC} shows the unbinned, folded \textit{K2} C3 long cadence light curve and the best fit \textsc{elc} model.

\begin{deluxetable}{lcc}[h!]
\tablewidth{0pt}
\tablecaption{Orbital and Apsidal Motion Parameters \label{orbpar}}
\tablehead{ \colhead{Parameter} & \colhead{Clausen (1991)} & \colhead{This Work}   }
\startdata  												
$P_s$ (days) 			& $  6.719695 	\pm	0.000003 $   	& $ 6.7196909  \pm 0.0000002 	$  \\ 
$T$ (BJD-2400000)    	& $  2444545.5215 \pm 0.0006$   	& $ 57165.3471\pm 0.0012	$  \\ 	
$e$                    		& $ 0.17 	\pm 0.01	$   			& $ 0.1758   	\pm 0.0012 	$  \\ 	
$\omega$ (deg)         	& $ 101.3 	\pm 0.7 	$   			& $ 103.04   	\pm 0.07 		$  \\ 	
$i$ (deg)              		& $ 88.4 	\pm 	0.1	$   			& $ 88.37  	\pm 0.04		$  \\ 
\\	 	
$K_1$ (km s$^{-1}$)    	& $ 84.2 	\pm 	0.5\tablenotemark{*}	$   	& $ 84.56   \pm 0.27 	$   \\	
$K_2$ (km s$^{-1}$)    	& $ 78.4 	\pm 	0.5\tablenotemark{*}	$   	& $ 77.83   \pm 0.27 	$   \\	
$\gamma$ (km s$^{-1}$)	& $ 9.9 	\pm 	0.3\tablenotemark{*}	$   	& $ 9.03   	 \pm 0.16 	$   \\	
$M_2 / M_1$           	 	& $ 1.07 	\pm 	0.009$   			& $ 1.087   	\pm 0.004	$   \\	
$a \sin i$ ($R_\odot$) 	& $ 21.28 	\pm 	0.14	$   			& $ 21.23	   	\pm 0.05 	$   \\ 	
\\ 	
$P_a$ (days)		  	& $ 6.719712 	\pm	0.000003	$   	& $6.719714	 \pm 0.000003	$  \\
$\dw$ (deg cycle$^{-1}$)	& $ 0.00090	\pm 	0.00010 	$   	& $0.00114	\pm 0.00017	$  \\
$U$ (years)			& $ 7400		\pm 	900		$   	& $5710 		\pm 830		$  \\
$\log \bar{k}_{2\ obs}$  	& $ -2.2		\pm 	0.1		$   	& $-2.04 		\pm 0.08		$  	
\enddata  
\tablenotetext{*}{Spectroscopic elements from \citet{imbert87}.}
\end{deluxetable}

There seems to be an artifact in the residuals during the secondary eclipse that we could not model, even after testing several limb darkening laws and coefficients from \citet{vanhamme93}. In the end, the logarithmic law produced the lowest residuals.   We also did not account for possible third light from a tertiary companion, which would dilute the eclipses and cause the inclination to be underestimated. However, the inclination is very close to $90^\circ$, so this effect would be small.

\subsection {Apsidal Motion}
The anomalistic period ($P_a$) of a binary system includes the apsidal motion ($\dw$) in the observed motion of the binary and is related to the sidereal period by $P_s = P_a (1 - \dw/360)$, where $\dw$ is in units of deg cycle$^{-1}$ \citep{hilditch}. $P_a$ and $\dw$ can be determined using eclipse timing, because apsidal motion causes the observed eclipse times to deviate from a linear ephemeris over time (often written as observed minus calculated time, $O-C$). One would calculate the predicted $O-C$ values for a given pair of $P_a$ and $\dw$, then compare these to the observed $O-C$ values in order to test different apsidal motion parameters. For example, \citet{kk86} first estimated $\dw = 0.0013 \pm 0.0001$ deg cycle$^{-1}$ for BW~Aqr, while \citet{clausen91} found $\dw = 0.0009 \pm 0.0001$ deg cycle$^{-1}$ using the method of \citet{ggp83}. \citet{lacy92} published a new method of determining the predicted $O-C$ values by using an iterative least-squares minimization to solve the apsidal motion equations numerically, without relying on the simplifications needed in previous studies. Using this method, \citet{bulut09} found $\dw = 0.00092 \pm 0.00015$ deg cycle$^{-1}$. 

\begin{figure}[h!]
\centering
\epsscale{1.2}
\plotone{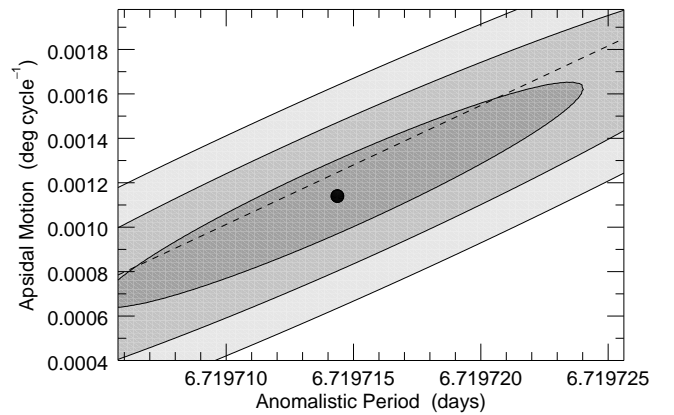}
\caption{$\chi^2$ contour map for various apsidal motion parameters. The best fit $\dw$ and $P_a$ are marked with the black circle, and the shaded regions represent the 1$\sigma$, 2$\sigma$, and 3$\sigma$ levels. The dashed line shows the correlation between $\dw$ and $P_a$ for our value of $P_s$. \label{lacycontour}} 
\end{figure}

\begin{figure*}
\centering
\epsscale{1.0}
\plotone{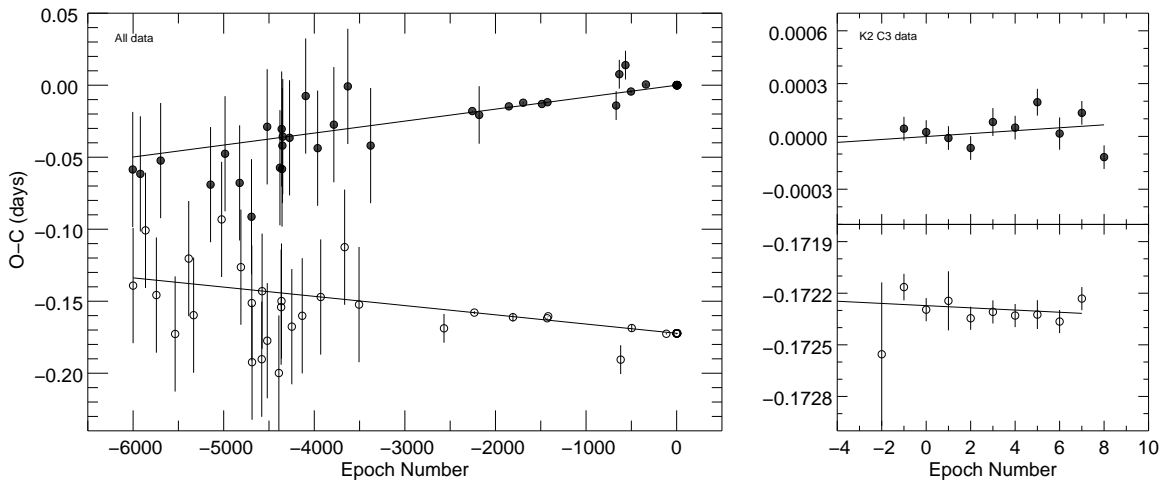}
\caption{Left: Ephemeris curve for all eclipse times as a function of the integer epoch number. The filled and open circles represent the primary and secondary eclipses, respectively, while the solid lines show the best fit model. Right: Portion of the ephemeris curve during the \textit{K2} C3 observations for the primary eclipses (top) and secondary eclipses (bottom).  \label{lacyOC}} 
\end{figure*}

With all of the literature eclipse times and the new \textit{K2} eclipse times in hand, we used the method of \citet{lacy92} to calculate the predicted $O-C$ for a grid of $P_a$ and $\dw$ values.  
While the apsidal motion parameters do depend on the eccentricity and are usually solved together, we held $e$ and $\omega$ fixed at the epoch of the \textit{K2} data because they are better constrained by the light curve through the eclipse durations and separations \citep{matson16}.  We also fixed the inclination and the time of primary eclipse to the values from the light curve solution. We calculated the $\chi^2$ for each pair of apsidal motion parameters to create the $\chi^2$ contour map shown in Figure~\ref{lacycontour}. The best fit apsidal motion parameters are $P_a = 6.719714 \pm 0.000003$ days and $\dw = 0.00114 \pm 0.00017$ deg cycle$^{-1}$, as listed in Table~\ref{orbpar}. The corresponding $O-C$ diagram is shown in Figure~\ref{lacyOC}. Our $\dw$ is slightly higher than that of \citet{clausen91}, whose errors are likely underestimated. 

Despite the eclipses measurements for BW~Aqr spanning over 100 years, the apsidal motion is so slow that the observations cover only a few percent of the total apsidal motion cycle. This slow precession and the correlation between $P_a$ and $\dw$ result in a family of solutions and large uncertainties in these parameters, as is evident in Figure~\ref{lacycontour}.  
Nonetheless, $P_a$ and $\dw$ can be used with the orbital parameters of an eccentric binary system to determine the system's internal structure constant as part of the apsidal motion test, which we discuss later in Section \ref{k2}.

% --------------------------------------------------------------------------------------------------------------------------------------
\section{Spectroscopic Analysis}\label{spectroscopy}

\subsection{Radial Velocities}
We used a two-dimensional cross-correlation algorithm \citep{matson16} to measure the radial velocities ($V_r$) of BW~Aqr. The template spectra for the primary and secondary components were created from \textsc{Bluered} model spectra of \citet{bluered} based on the effective temperatures, surface gravities, and rotational velocities from \citet{clausen91}.  \citet{clausen10} completed an abundance analysis for BW~Aqr and found [Fe/H] $= -0.07 \pm 0.11$, so we interpolated the \textsc{Bluered} models to $\log Z/Z_\sun=-0.07$ for our analysis.  Note that \citet{clausen10} used the solar abundances from \citet{grevesse07} with $Z_\odot = 0.012$, while the \textsc{Bluered} models use the solar abundances from \citet{ag89} with $Z_\odot = 0.0189$. 

The cross-correlation algorithm first determines the best fit velocity separation of the components; the model for the primary star is combined with models of the secondary shifted by different relative velocities, and the maximum correlation is recorded for each trial velocity. A parabolic fit to the resulting maximum correlations determines the relative velocity of the secondary star with respect to the primary.  The template for this separation is then cross-correlated with the observed spectrum to determine the absolute velocities of each component.

This process was repeated for all echelle orders covering $4500-7000$\AA, discarding any extreme outliers that were more than 3$\sigma$ away from the mean or where the primary and secondary components' identities were interchanged. The radial velocities for each night were calculated using a weighted average of the velocities from the remaining echelle orders, and the errors correspond to the standard deviations of these values. The final radial velocities, re-derived with templates formed from the updated atmospheric parameters from Section \ref{atmos}, are listed in Table~\ref{rvtable}.

\begin{deluxetable*}{cccccccc}
\tablewidth{\textwidth}
\tabletypesize{\normalsize}
\tablecaption{Radial Velocity Measurements \label{rvtable}}
\tablehead{ \colhead{Date} & \colhead{Orbital} & \colhead{$V_{r1}$} & \colhead{$\sigma_1$} & \colhead{Residual}  & \colhead{$V_{r2}$} & \colhead{$\sigma_2$} & \colhead{Residual}   \\  
\colhead{(HJD-2400000)} & \colhead{Phase} & \colhead{(km~s$^{-1}$)} & \colhead{(km~s$^{-1}$)} & \colhead{(km~s$^{-1}$)} & \colhead{(km~s$^{-1}$)} & \colhead{(km~s$^{-1}$)} & \colhead{(km~s$^{-1}$)}}
\startdata 
57162.864    &  0.630    &  $ 68.16$    &  0.99  &   $ 0.36$   &  	$ -44.80$   &  $ 0.44$   	&  $  0.26$  \\ 
57163.834    &  0.775    &  $ 89.94$    &  0.70  &   $-0.21$   &  	$ -65.18$   &  $ 0.78$   	&  $  0.45$  \\
57164.886    &  0.931    &  $ 42.00$    &  0.71  &   $ 4.60$   &  	$ -19.63$   &  $ 0.58$   	&  $ -2.55$  \\
57166.856    &  0.224    &  $-70.65$    &  0.77  &   $ 0.88$   &  	$  83.47$   &  $ 1.11$    	&  $  0.30$  \\ 
57168.751    &  0.506    &  $ 32.38$    &  1.04  &   $ 5.28$   &  	$ -10.93$   &  $ 1.25$   	&  $ -3.32$  \\
57169.840    &  0.669    &  $ 78.59$    &  0.75  &   $ 1.16$   &   	$ -53.16$   &  $ 0.53$  	&  $  0.77$  \\ 
57170.816    &  0.814    &  $ 87.64$    &  0.58  &   $ 0.80$   &   	$ -62.08$   &  $ 1.14$  	&  $  0.51$  \\ 
57171.784    &  0.958    &  $ 13.79$    &  8.15  &   $-4.62$   &   	$    7.52$   &  $ 3.08$   	&  $  7.13$  \\ 
57172.928    &  0.128    &  $-73.89$    &  0.62  &   $ 2.51$   &   	$  86.81$   &  $ 0.60$   	&  $ -0.84$  \\
57174.902    &  0.422    &  $-10.05$    &  0.83  &   $-5.21$   &   	$  26.62$   &  $ 0.56$   	&  $  4.84$  \\
57175.619    &  0.528    &  $ 45.57$    &  0.99  &   $10.55$  &   	$ -21.87$   &  $ 0.41$   	&  $ -6.96$  \\
57176.822    &  0.707    &  $ 87.17$    &  0.93  &   $ 2.23$   &   	$ -61.42$   &  $ 0.54$   	&  $ -0.58$  \\
57177.757    &  0.847    &  $ 77.47$    &  0.72  &   $-1.84$   &   	$ -53.09$   &  $ 0.45$   	&  $  2.58$  \\
57193.747    &  0.226    &  $-70.28$    &  0.72  &   $ 0.88$   &   	$  82.90$   &  $ 0.82$   	&  $  0.08$ 
\enddata 
\end{deluxetable*}

\begin{figure}[h!]
\centering
\epsscale{1.2}
\plotone{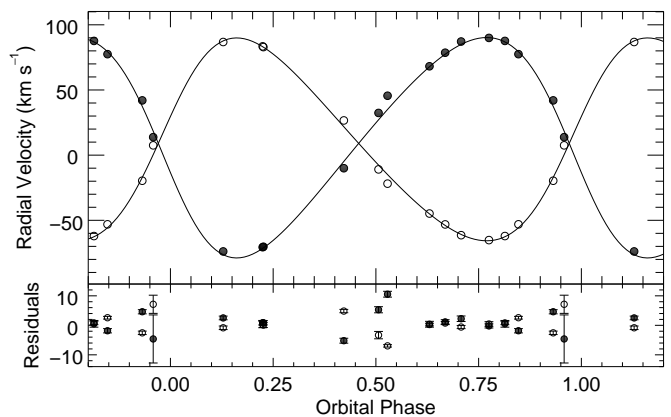}
\caption{Radial velocity curve and residuals for BW~Aqr, phased relative to the epoch of periastron. The filled circles represent the primary star and the open circles represent the secondary star, while the solid lines correspond to the best fit models. \label{rvcurve}} 
\end{figure}

\subsection{Orbital Parameters \label{specorbpar}}
We used the orbit fitting code \textsc{rvfit}\footnote{\href{http://www.cefca.es/people/~riglesias/rvfit.html}{www.cefca.es/people/$\sim$riglesias/rvfit.html}} by \citet{rvfit} to determine the orbital parameters of the system. This code uses adaptive simulated annealing to fit for any combination of the orbital period ($P$), epoch of periastron ($T$), eccentricity ($e$), longitude of periastron ($\omega$) of the primary star, systemic velocity ($\gamma$), and velocity semi-amplitudes ($K_1$, $K_2$).  Two of our spectra were taken during an eclipse (at phases $\phi=0.422, 0.958$), so the radial velocities measured at these times are not accurate. Unfortunately, these points correspond to phases which  influence the shape of the radial velocity curve and resulting fits for $e$ and $\omega$. These parameters, as well as the orbital period and epoch of periastron, are much better constrained by the light curve model, so we held $P$, $T$, $e$ and $\omega$ fixed and fit only for $\gamma$, $K_1$, and $K_2$.  The best fit values are listed in Table~\ref{orbpar}, along with other derived quantities such as the mass ratio ($q=M_2/M_1$) and the projected semi-major axis ($a \sin i$). We calculated the errors on the three fitted parameters using the Monte Carlo Markov Chain (MCMC) feature of \textsc{rvfit}, where the error in each parameter corresponds to the standard deviation of the Gaussian fit to each MCMC result. These errors are roughly $\sqrt{N}$ better than the individual errors in radial velocity, as expected. The best fit model radial velocity curve is shown in Figure~\ref{rvcurve} and the residuals are listed in Table~\ref{rvtable}.  The systemic velocity is very similar to that of \citet{imbert87} and we found no periodicity in the radial velocity residuals, which likely excludes the presence of a nearby tertiary companion.

\break

\subsection{Atmospheric Parameters \label{atmos}}
In order to determine the atmospheric parameters for each component of BW~Aqr, we used the Doppler tomography algorithm of \citet{tomography} to reconstruct the individual component spectra. We adopted \textsc{Bluered} template spectra of each component with atmospheric parameters from \citet{clausen91} as starting estimates in the algorithm. Doppler tomography also requires an input for the flux ratio ($F_2/F_1$), which determines the strengths of the absorption lines in the reconstructed spectra and affects any resulting fits for the effective temperature ($\teff$). 

In order to determine the flux ratio of BW~Aqr, we created a $\chi^2$ contour across a grid of flux contributions from each component ($F_1$, $F_2$) as follows. The Balmer lines are the absorption lines most sensitive to temperature for F-type stars, so we ran Doppler tomography on the H$\alpha$ echelle order to create reconstructed spectra of each component for our grid of $F_1$ and $F_2$ values.  At each grid point, we fit for $\teff$ using \textsc{mpfit} \citep{markwardt} to compare \textsc{Bluered} models of various $\teff$ to the reconstructed spectra of the H$\alpha$ order. We then used the best fit model spectrum to calculate $\chi^2$ across the order, creating a $\chi^2$ contour for our grid of flux values. We fit a parabola to the projected $\chi^2$ curve in each dimension to determine the best flux contributions to be  $F_1 = 0.36 \pm 0.03$ and $F_2 = 0.54 \pm 0.04$, corresponding to a flux ratio of $F_2/F_1 = 1.45 \pm 0.08$ at 6563\AA. 

\begin{figure}[h!]
\centering
\epsscale{1.2}
\plotone{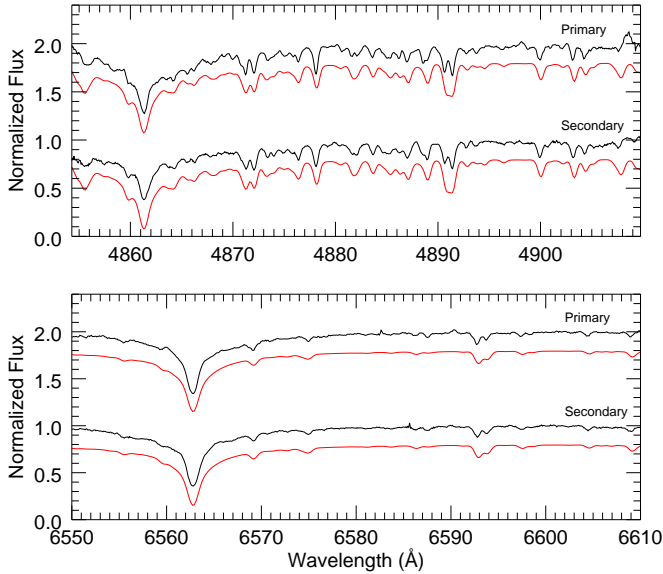}
\caption{Reconstructed spectra of BW~Aqr for the H$\beta$ (top) and H$\alpha$ (bottom) echelle orders. The reconstructed spectra are shown in black, and the best fit model spectra are shown in red and offset by -0.2 flux units. \label{recspec}} 
\end{figure}

We created new reconstructed spectra with this flux ratio to use in determining the final effective temperatures of BW~Aqr. Because $F_1 + F_2 \le 1$, we think that there is some extra background flux that was not removed during reduction process. To avoid the issue of unreliable absolute line depths in the resulting temperatures fits, we instead used line ratios. We chose eight pairs of absorption lines with varying dependencies on temperature and well defined continuum levels, which compared an \ion{Fe}{1} line to either an \ion{Fe}{2} line or the core of H$\alpha$. For each pair of absorption lines, we measured the line depth ratios in the reconstructed spectra and model spectra of various $\teff$, then interpolated between the model line ratios to find the effective temperature corresponding to the observed ratio. We took the mean and standard deviation of the results from all line pairs to calculate the final effective temperatures for BW~Aqr, $T_{\rm eff \ 1} = 6370 \pm 270$ K and $T_{\rm eff \ 2} = 6320 \pm 220$ K.  The temperature ratio is then $T_{\rm{eff\ 2}} / T_{\rm{eff\ 1}}=0.992 \pm 0.090$, which is very similar to the ratio from the light curve analysis. These effective temperatures correspond to spectral types of about F6 and F7 using the temperatures from \citet{gray}. Our results are also consistent with the results from \citet{clausen10}, though they achieved smaller errors from their abundance analysis. Example reconstructed spectra of BW~Aqr for the H$\alpha$ and H$\beta$ echelle orders are shown in Figure \ref{recspec}, along with the final, best fit \textsc{Bluered} model templates.  

Finally, we determined the projected rotational velocity ($V \sin i$) of each component from the individual, reconstructed spectra. We identified twelve strong, well separated metal absorption lines to use in comparing the reconstructed spectra to \textsc{Bluered} model spectra of varying $V \sin i$. For each absorption line, we calculated $\chi^2$ of the models and fit a parabola to the curve to find the best fit $V \sin i$. We calculated the 1$\sigma$ errors from the velocities corresponding to $\chi^2 \leq \chi^2_{min} + 1$. The final $V \sin i$ for each component were then calculated from the weighted averages of the results from all twelve absorption lines, which we found to be 
$V_1 \sin i = 12.6~\pm~1.9$~km~s$^{-1}$ and 
$V_2 \sin i = 14.6~\pm~2.0$~km~s$^{-1}$.  
Both components are rotating near the projected synchronous velocities of 13.4~km~s$^{-1}$ and 15.4~km~s$^{-1}$, which is rather common, as \citet{lurie17} found that 72\% of \textit{Kepler} eclipsing binaries with orbital periods between 2 and 10 days are rotating at the synchronous velocities.

% --------------------------------------------------------------------------------------------------------------------------------------
\section{Results}\label{results}

\subsection{Absolute Parameters}
Combining the results from the spectroscopic and light curve analyses, the absolute parameters for BW~Aqr are 
$M_1 = 1.365 \pm 0.008 \ M_\sun$, 
$M_2 = 1.483 \pm 0.009 \ M_\sun$,  
$R_1  = 1.782 \pm 0.021 \ R_\sun$, and 
$R_2  = 2.053 \pm 0.020 \ R_\sun$. 
The surface gravities of each component are then $\log g_1 = 4.071 \pm 0.010$ and $\log g_2 =  3.985 \pm 0.009$.  A summary of our results are listed in Table~\ref{atmospar} and have errors in mass and radius of about $0.6\%$ and $1\%$, respectively. They are also consistent with the results of \citet{clausen91}.

\begin{deluxetable}{lcc}
\tablewidth{0pt}
\tablecaption{Astrophysical Parameters \label{atmospar}}
\tablehead{ \colhead{Parameter} & \colhead{Primary} & \colhead{Secondary} }
\startdata					
Mass ($M_\sun$)			& 1.365 $\pm$ 0.008		& 1.483 $\pm$ 0.009		 \\
Radius  ($R_\sun$) 			& 1.782 $\pm$ 0.021		& 2.053 $\pm$	0.020	\\ 
$\teff$ (K)					& 6370  $\pm$   270		& 6320  $\pm$   220		\\ 
$\log g$ (cgs)				& 4.071 $\pm$ 0.010		& 3.985 $\pm$ 0.009		\\ 
 $V \sin i$ (km~s$^{-1}$)		& 12.63 $\pm$ 1.86		& 14.56 $\pm$ 2.01		\\ 
$[\rm{Fe/H}]$ 	  			& \multicolumn{2}{c}{-0.07 $\pm$ 0.11\tablenotemark{*}}	\\
 $F_2/F_1$  (at 6563\AA)		& \multicolumn{2}{c}{ $1.45 \pm 0.08$ } 			
\enddata
\tablenotetext{*}{Fixed from \citet{clausen10}}
\end{deluxetable}

\subsection{Comparison with Evolutionary Models}
We compared the observed mass and radius to the predictions of several stellar evolution models: the Yonsei-Yale $Y^2$ models of \citet{y2}, the Geneva models of \citet{geneva},  the Granada models of \citet{claret04, claret06}, the MESA code of \citet{mesa, mesa4}, and the Victoria-Regina models of \citet{vandenberg06}.  Non-rotating models with scaled solar abundance were used throughout. Note that each model uses a different solar metallicity prescription, which causes a slight scatter in the zero age main sequence positions of each mass track. For each evolutionary model, we estimated the ages of each component star by interpolating the age of the evolutionary tracks at the observed radii. 

\begin{figure*}[t!]
\centering
\epsscale{1.0}
\plotone{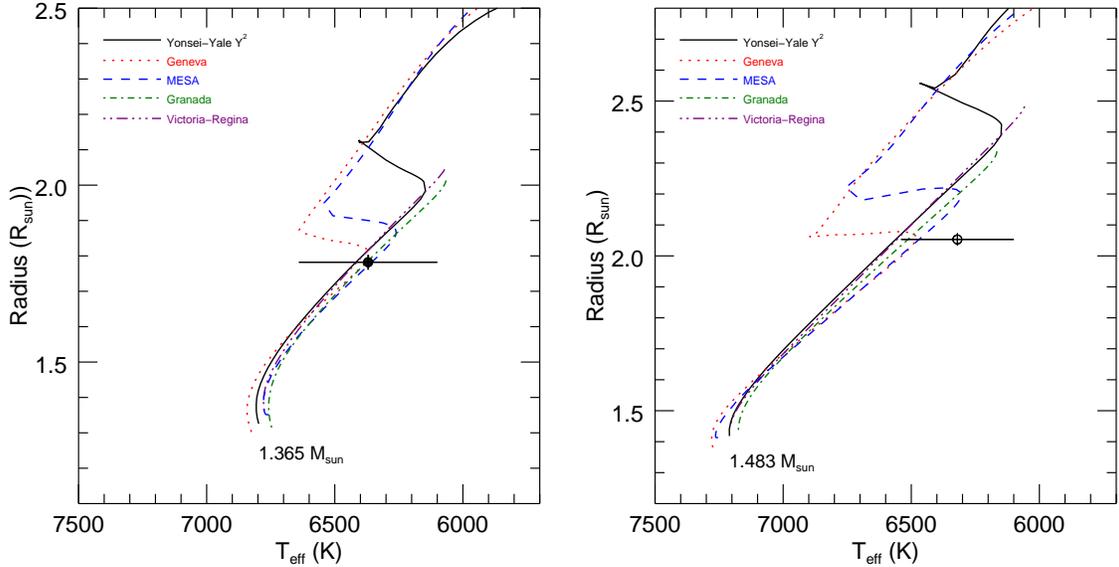}
\caption{Evolutionary tracks for the primary star (left) and secondary star (right) for [Fe/H]$=-0.07$. Full mass tracks for the $Y^2$ models (black, solid lines), Geneva models (red, dotted lines), and MESA models (blue, dashed lines) are shown. The main sequence portions of the Granada models (green, dot-dashed lines) and the Victoria-Regina models (purple, dot-dot-dashed) are also shown.  \label{evo} }
\end{figure*}

\begin{figure}[t!]
\centering
\epsscale{1.2}
\plotone{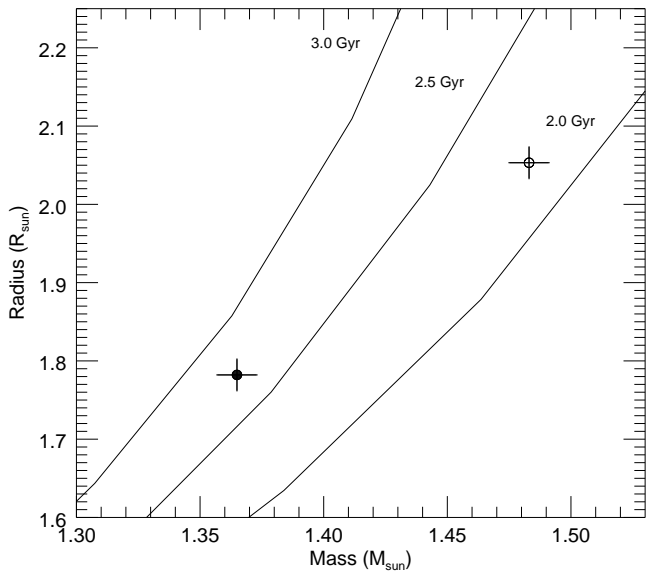}  
\caption{Yonsei-Yale $Y^2$ isochrones for [Fe/H]$=-0.07$ and ages of 2.0, 2.5, and 3.0 Gyr. The positions of the primary and secondary components of BW~Aqr are shown as the filled and open circles. \label{iso} }
\end{figure}

For the Yonsei-Yale $Y^2$ models\footnote{\href{http://www.astro.yale.edu/demarque/yystar.html}{astro.yale.edu/demarque/yystar.html}}, we used the evolutionary track interpolator provided to create tracks at the observed masses and metallicity, shown in Figure~\ref{evo}. These models use the step-function method to characterize convective core overshooting based on an overshoot parameter $\Lambda_{\rm ov}$ (sometimes written as $\alpha_{\rm ov}$). The amount of overshooting therefore corresponds to $d = \Lambda_{\rm ov} H_p$, where $H_p$ is the pressure scale height at the convective boundary \citep{y2}.  In the $Y^2$ models, $\Lambda_{\rm ov}$ is a function of mass and metallicity and is about 0.13 and 0.20 for the components of BW~Aqr. These models predict ages of 2.72 Gyr and 2.19 Gyr for the primary and secondary components and a mean age of 2.45 Gyr. The difference in age between each component is about 21\% of the mean age and is more easily seen in the $Y^2$ isochrones shown in Figure~\ref{iso}.

For the Geneva models, we used the online model interpolator\footnote{\href{http://obswww.unige.ch/Recherche/evoldb/index/Interpolation}{obswww.unige.ch/Recherche/evoldb/index/Interpolation}} to produce the evolutionary tracks shown in Figure~\ref{evo}.  The Geneva code also uses the step-function method to increase $\alpha_{\rm ov}$ with mass, but with a slower increase than the $Y^2$ models. For both components of BW~Aqr, $\alpha_{\rm ov}= 0.05$.   These models predict ages of 2.80 Gyr and 2.20 Gyr and a mean age of 2.50 Gyr  (24\% difference in age).

The Granada models of \citet{claret04, claret06}\footnote{\href{http://vizier.u-strasbg.fr/viz-bin/VizieR?-source=J/A\%2BA/424/919}{vizier.u-strasbg.fr/viz-bin/VizieR?-source=J/A\%2BA/424/919}} cover a grid of mass and metallicity values with a fixed overshoot parameter, $\alpha_{\rm ov} = 0.20$. We interpolated between models to the observed mass and metallicity values, but the red hook is very difficult to interpolate across so only the main sequence parts of the evolutionary tracks are shown in Figure~\ref{evo}.  These models predict ages of 2.87 Gyr and 2.22 Gyr with a mean age of 2.55 Gyr (26\% age difference). 

The MESA code\footnote{\href{http://www.mesa.sourceforge.net}{mesa.sourceforge.net}} computes models for any given mass and metallicity, which the user can specify in the input files. To characterize convective core overshooting, MESA uses the diffusion method based on the overshoot parameter, $f_{\rm ov}$, which the user can also specify. \citet{ct17} created a semi-empirical calibration of $f_{\rm ov}$ based on well-studied eclipsing binaries, so we estimated $f_{\rm ov} = 0.002$ and $0.008$ for BW~Aqr from their calibration. Additionally, MESA can employ the solar abundance prescriptions from several different sources, so we chose to use the solar abundances from \citet{ag89} to match the \textsc{Bluered} models. The MESA models predict ages of 2.31 Gyr and 1.89 Gyr and a mean age of 2.10 Gyr (20\% age difference).  

The Victoria-Regina models\footnote{\href{http://www.cadc-ccda.hia-iha.nrc-cnrc.gc.ca/community/VictoriaReginaModels/}{cadc-ccda.hia-iha.nrc-cnrc.gc.ca/community/VictoriaReginaModels/}} also cover a grid of mass and metallicity values, so we interpolated between models to the observed values of mass and metallicity, using only the main sequence portions of the evolutionary tracks. The Victoria-Regina models use a different implementation of convective overshooting, based on the Roxburgh criterion \citep{roxburgh78, roxburgh89}. The free parameter $F_{ov}$ is a function of mass and metallicity, which corresponds to $F_{ov} \sim$ 0.2 and 0.4 for BW~Aqr. These models predict ages of $2.50$ Gyr and $2.29$ Gyr with a mean age of 2.30 Gyr (18\% age difference).

As seen in Figure~\ref{evo}, all evolutionary models are able to fit the observed temperatures and radii individually, but none of the models are able to reproduce the observed properties with a single isochrone. There seems to be a difference of about 20\% between the ages of the primary and secondary stars, with the secondary star predicted to be younger in all cases. This is most easily seen for the $Y^2$ isochrones shown in Figure~\ref{iso}, where the observed slope between components is shallower than the slope of the isochrones.

\break

\subsection{Internal Structure Constant}\label{k2}
Absolute stellar parameters can be combined with the apsidal motion parameters to probe the internal structure of a binary system because the rate of apsidal motion depends on the internal mass distributions of the component stars. However, one cannot calculate the contributions of each star's internal structure to the observed apsidal motion individually; one can calculate only a mean observed internal structure constant ($\log \bar{k}_{2\ obs}$) to compare to stellar evolution theory as part of the apsidal motion test \citep{cg10}.  

The observed apsidal motion rate has a classical contribution ($\dw_{clas}$) and a relativistic contribution ($\dw_{rel}$). The relativistic component can be calculated from the orbital elements and component masses as described in Appendix~A. We found $\dw_{rel} = 0.00032(5)$ for BW~Aqr, which constitutes about 27\% of the observed apsidal motion rate. Once the total observed and relativistic apsidal motion rates are known, one can calculate the classical apsidal motion rate and the corresponding $\log \bar{k}_{2\ obs}$ value using the equations in Appendix~A. Using this method, we found $\log \bar{k}_{2\ obs} =  -2.02 \pm 0.08$ for BW~Aqr. 

From certain evolutionary models, one can predict $\log k_{2}$ for each star in the binary system ($\log k_{21\ theo}$, $\log k_{22\ theo}$). Because rapidly rotating stars are more centrally condensed and have lower $\log k_{2}$ than slowly rotating stars, these theoretical $\log k_{2}$ values must be corrected for rotation \citep{claret99} using the equations in Appendix~B. For BW~Aqr, the corrections due to rotation are small, $\Delta \log k_{21} =-0.00073 \pm 0.00011$ and $\Delta \log k_{22} =-0.00097\pm 0.00015$.  Additionally, time-dependent tidal distortions due to non-synchronous rotation affect the predicted apsidal motion rate. \citet{claret02} calculated the necessary corrections for several binary systems and found $\Delta_{dyn}=0.00054$ for BW~Aqr.  One can then calculate an average theoretical internal structure constant ($\log \bar{k}_{2\ theo}$) using the equations described in Appendix~B to compare with the value from observations.

We calculated $\log \bar{k}_{2\ theo}$ from both the Granada and MESA evolutionary models.
The Granada models provide theoretical $\log k_{2}$ values for a grid of masses, $\log g$, and [Fe/H], so we interpolated between these values to determine $\log k_{21\ theo}=-2.30$ and $\log k_{22\ theo}=-2.42$ for each component of BW~Aqr.  After correcting for rotation and dynamic tides, taking the weighted average and then the logarithm, we found $\log \bar{k}_{2\ theo} =  -2.37$ for the Granada models.    For the MESA models, $\log k_{2}$ is not a direct output, but can be calculated from the density and interior mass profiles. Using the procedure in Section 1 of \citet{cisneros70} and described in Appendix~C, we integrated the Radau equation to calculate $\log k_{21\ theo}= -2.27$ and $\log k_{22\ theo}=-2.36$. We corrected for rotation and dynamic tides, took the weighted average and logarithm, and found $\log \bar{k}_{2\ theo} =  -2.33$ for the MESA models. 

Both sets of models predict lower $\log k_{2}$ than the observed value. However, the first condition in the apsidal motion test is that the models must be able to match the observed surface properties of the stars \textit{at the same age}, because the internal structure constant is highly dependent on the radius of the stars. Neither the Granada nor MESA models were able to fit the temperatures and radii of both components with a single isochrone, so it is not surprising that the theoretical internal structure constants do not agree with the observed value.

% --------------------------------------------------------------------------------------------------------------------------------------
\section{Discussion}\label{discussion}

We determined the fundamental parameters for the F-type eclipsing binary, BW~Aqr, to within $0.6\%$ for mass and $1.2\%$ in radius.  We compared these results to several stellar evolution models, none of which could fit both components of BW~Aqr at the same age (to within $\sim 5\%$).  All models predicted the more massive component to be younger than the less massive component by $19-26$\%. \citet{clausen10} also noted this age discrepancy in BW~Aqr, even with the slightly larger errors in radius.

One possible explanation is that the observed radii need revision. The secondary star is much smaller than the models predict at the observed temperature by at least 4$\sigma$. Totally eclipsing systems allow us to determine $R_1/a$ and $R_2/a$ very accurately, but partially eclipsing systems only allow us to constrain $(R_1+R_2)/a$ to such accuracy. This creates a valley of solutions: as one star is made smaller, the other would be made larger and create a similarly good fit to the light curve. However, we did take this into account in the \textsc{elc} fit and error budget. 

 A similar age discrepancy has been found in other evolved, F-type eclipsing binaries: GX Gem, BK Peg, V442 Cyg \citep{clausen10}, CO And  \citep{lacy10}, BF Dra  \citep{lacy12}, and AQ Ser \citet{torres14}. The more massive component was found to be younger in all of these systems, suggesting that the age discrepancy is not due to observational error.  \citet{torres14} tested different core overshooting and mixing length parameters for AQ Ser but could not resolve the age discrepancy. They postulated a dependence of the overshooting parameter on evolutionary state, in addition to the current dependence on mass and metallicity\footnote{While this paper was in review, \citet{ct18} published updated MESA models of these seven systems. By fine tuning both the core overshooting and mixing length parameters for each component individually, they were able to fit the observed properties of both components to within a 5\% age difference for all systems except AQ Ser.}.  

We also completed an apsidal motion analysis for the system and calculated the mean observed internal structure constant. We found that the observed value is larger than the theoretical predictions of both the Granada and MESA models, implying that the BW~Aqr stars are less centrally condensed than predicted by models. This is likely due to the failure of the models to produce the observed surface parameters of BW~Aqr at the same age. There has been some disagreement in the past between the observed and theoretical $\log \bar{k}_2$ values, but correcting for rotation, relativistic effects, core overshooting, and dynamic tides has significantly reduced the disagreement. Furthermore, \citet{cg10} compiled well studied binaries with mass and radius errors less than 2\% and found that the observed $\log \bar{k}_2$ values matched the theoretical predictions within the errors.

These disagreements between the observations and theory are interesting problems with far reaching consequences.  F-type stars lie in the mass range where stars begin to develop convective cores ($1.1-1.7 M_\odot$). F-type eclipsing binaries, especially those with evolved components, are used to calibrate the treatment of convective core overshooting in evolutionary models \citep{ct16, ct17}. This has larger implications for determining the ages of single stars, exoplanet properties, and the star formation history of the galaxy. Therefore, it would be quite beneficial to study other evolved, F-type eclipsing binaries to solve these discrepancies.

% --------------------------------------------------------------------------------------------------------------------------------------
\acknowledgments
We would like to thank Zhao Guo for his help with the MESA code, as well as the CTIO staff for taking the CHIRON observations. The CTIO 1.5 m telescope is operated by the SMARTS Consortium. This paper includes data collected by the Kepler mission, which was competitively selected as the tenth Discovery mission. Funding for this mission is provided by NASA's Science Mission Directorate. \textit{K2} data were obtained from the Mikulski Archive for Space Telescopes (MAST). STScI is operated by the Association of Universities for Research in Astronomy, Inc., under NASA contract NAS5-26555. Support for MAST for non-HST data is provided by the NASA Office of Space Science via grant NNX09AF08G and by other grants and contracts.  This work also made use of PyKE (Still \& Barclay 2012), a software package for the reduction and analysis of Kepler data. This open source software project is developed and distributed by the NASA Kepler Guest Observer Office. This material is based upon work supported by the National Science Foundation under grant No.~AST-1411654. Institutional support has been provided from the GSU College of Arts and Sciences and from the Research Program Enhancement fund of the Board of Regents of the University System of Georgia, administered through the GSU Office of the Vice President for Research and Economic Development.

{\it Facilities: Kepler/K2, CTIO:1.5m}

\clearpage

% --------------------------------------------------------------------------------------------------------------------------------------
\begin{appendix}

\section{Observed Internal Structure Constant \label{apA}}
This section details how to calculate the observed internal structure constant from the apsidal motion and orbital parameters using the procedure in Section 5.1 of \citet{claret02}. The observed apsidal motion rate has classical and relativistic contributions, where $$\dw_{obs} = \dw_{clas} + \dw_{rel}.$$ All apsidal motion rates are expressed in units of deg cycle$^{-1}$.
The relativistic component can be calculated using Eq. 3 from \citet{gimenez85}, $$\dw_{rel} = 5.45 \times 10^{-4}\ \frac{1}{(1-e^2)} \Bigg(\frac{m_1+m_2}{P_s}\Bigg)^{2/3}$$ where $P_s$ is the sidereal period in days, $e$ is the orbital eccentricity, and $m_1$ and $m_2$ are the stellar masses in units of $M_\odot$.  Because only the classical component holds information about the internal structure of the stars, one must now isolate $\dw_{clas}$ using, $$ \dw_{clas} = \dw_{obs} - \dw_{rel}.$$

The classical component has contributions from both stars in the binary system, which cannot be calculated individually. We can only calculate a mean observed internal structure constant ($\log \bar{k}_{2\ obs}$) for the system to compare to theory.
First, calculate the contribution weights for each star,
$$c_{21} = \Bigg[ \bigg(\frac{\Omega_1}{\Omega_{K_1}}\bigg)^2 \bigg(1 + \frac{m_2}{m_1}\bigg)f(e) + \frac{15\ m_2}{m_1}g(e) \Bigg] \ \Bigg(\frac{R_1}{a}\Bigg)^5 $$
$$c_{22} = \Bigg[ \bigg(\frac{\Omega_2}{\Omega_{K_2}}\bigg)^2 \bigg(1 + \frac{m_1}{m_2}\bigg)f(e) + \frac{15\ m_1}{m_2}g(e) \Bigg] \ \Bigg(\frac{R_2}{a}\Bigg)^5 .$$
Here, $\Omega_i/\Omega_K$ is the ratio of the observed rotational velocity of star $i$ to the synchronous velocity, and $f(e)$ and $g(e)$ are functions of the eccentricity,
$$f(e) = (1-e^2)^{-2}$$		$$g(e) = \frac{(8+12e^2 + e^4)\ f(e)^{2.5}}{8} .$$

Then the mean observed internal structure constant can be calculated using
$$ \bar{k}_{2\ obs} = \frac{1}{c_{21}+c_{22}}\ \frac{P_s}{U_{clas}} = \frac{1}{c_{21}+c_{22}} \ \frac{\dw_{clas}}{360}$$
where $\dw_{clas}$ is in units of deg cycle$^{-1}$. The internal structure constant is often written as $\log \bar{k}_{2\ obs}$. 
To calculate the error in $\bar{k}_{2\ obs}$, we propagated the errors in  $c_{21}$, $c_{22}$, and $\dw_{clas}$ analytically through the above equations. Because $c_{21}$ and  $c_{22}$ depend strongly  on the relative radii as $(R/a)^5$, we assumed that the only source of error in $c_{21}$ and $c_{22}$ were the errors in $R/a$.

\vspace{1cm}

\section{Theoretical Internal Structure Constant \label{apB}}
From certain evolutionary models, we can predict $\log k_{2}$ for each star in the binary system ($\log k_{21\ theo}$ and $\log k_{22\ theo}$). For example, the Granada models provide $\log k_2$ values directly for a grid of masses, surface gravities, and metallicities, so we can interpolate $\log k_2$ from the observed properties. The MESA models do not output $\log k_2$ directly, but it can be calculated from other outputs using the process described Appendix~C below.  
We then need to correct  $\log k_{21\ theo}$ and $\log k_{22\ theo}$ for the effects of rotation and dynamic tides, as follows.

We calculate the rotational correction using the equations from \citet{claret99},
 $$\log {k}_{2i\ theo}\ [corrected] =  \log k_{2i\ theo}  - \lambda_i $$
 $$\lambda_i  = \frac{2 V_i^2}{3g_iR_i}$$  
 where $V_i$ is the rotational velocity, $g_i$ is the surface gravity, and $R_i$ is the radius of each component.  
 
We correct for the effects of dynamic tides using Eq. 17 from \citet{claret02},
$$\Delta_{dyn} =  \frac{k_{2}  - k_{2\ dyn}}{k_{2\ dyn}}$$
where $k_{2}$ is the uncorrected value and $k_{2\ dyn}$ is the corrected value that includes the effects of dynamic tides. $\Delta_{dyn}$ can only be found analytically and is listed for several binary systems in Table 3 of  \citet{claret02}.  We then solve for the corrected value of $k_2$, 
$$k_{2\ dyn}  = \frac{k_{2}}{   1 + \Delta_{dyn} }   \hspace{0.5cm} \to \hspace{0.5cm} k_{2i\ theo}\ [corrected] =  \frac{k_{2i\ theo}}{   1 + \Delta_{dyn} }  .$$

Finally, the mean theoretical internal structure constant can be calculated from the corrected, individual values with
$$ \bar{k}_{2\ theo} = \frac{c_{21}\ k_{21\ theo}\  +\  c_{22}\ k_{22\ theo}} {c_{21}+c_{22}} $$
where $c_{21}$ and $c_{21}$ are the same weighting contributions as given in Appendix~A.

\vspace{1cm}

\section{Using MESA models \label{apC}}
This section details the process of calculating $\log k_{2 \ theo}$ from a MESA model using the method of \citet{cisneros70}. We create a MESA model for each component star, calculate the individual $\log k_{2}$ values, and then calculate the weighted average.

MESA outputs the density and interior mass profiles in the \texttt{profile\#\#.data} files for each age step, so we adopt the file corresponding to the age of binary system. The variables needed are:
\begin{itemize}
\item[ ]  $r$ =  the distance from center of star in $R_\odot$
\item[ ] $m(r)$ =  the mass interior to $r$ in $M_\odot$
\item[ ] $\rho(r)$ = the density at radius $r$ in cgs units.
\end{itemize}
From these variables, one can calculate:
\begin{itemize} 
\item[ ] $\bar{\rho}(r)$ = the mean density of the sphere interior to $r$
\item[ ] $\frac{\rho(r)}{\bar{\rho}(r)}$ = the density ratio
\item[ ] $\frac{d}{dr}\Big(\frac{\rho}{\bar{\rho}}\Big)$ =  the derivative of the density ratio.
\end{itemize}

The next step is to integrate the Radau Equation,
$$r \frac{dy_j(r)}{dr} + 6\ \frac{\rho(r)}{\bar{\rho}(r)}\ \Big(y_j(r) + 1\Big) + y_j(r)\Big(y_j(r) -1\Big) = j\big(j+1\big)$$
using the Runge-Kutta method to solve for $y_j(r)$ iteratively, working from the center outwards. The function $y_j(r)$ is a measure of the deviation from sphericity in the orders $j=2,3,4$ that correspond to $k_2, k_3, k_4$ \citep{claret02}. The structure constant of interest is $k_2$, so only $j=2$ is used.  The central boundary conditions needed for the first iteration are:
\begin{itemize}
\item[ ] $r(0) = 0$ at center of the star
\item[ ] $y(0) = j-2 = 0$, from solving the Radau equation at $r=0$ and taking the only positive (physical) solution. 
\item[ ] From \citet{poincare}, 
	$$\frac{dy_j}{dr}\Big|_0 = \frac{-3(j-1)}{j+1}\ \frac{d}{dr}\Bigg(\frac{\rho}{\bar{\rho}}\Bigg)\Bigg|_0 .$$
\end{itemize}

Then, one can calculate $k_2$ from $y_2(r)$ using, 
$$ k_j = \frac{j+1-y_j(R)}{2 (j+y_j(R))} \hspace{0.5cm} \to \hspace{0.5cm} k_{2\ theo} = \frac{3-y_2(R)}{4 + 2\ y_2(R)}$$
where $y_j(R)$ is the value at the surface. 

We repeat this process for each component star in order to determine $\log k_{21\ theo}$ and $\log k_{22\ theo}$ and then correct for the effects of dynamic tides and rotation as described above. Finally, one can calculate the mean theoretical value ($\bar{k}_{2\ theo}$) using the same weighting factors as given in Appendix~A,
$$ \bar{k}_{2\ theo} = \frac{c_{21}\ k_{21\ theo}\  +\  c_{22}\ k_{22\ theo}} {c_{21}+c_{22}} $$ and take the logarithm to arrive at $\log \bar{k}_{2\ theo}$.

\end{appendix}

% --------------------------------------------------------------------------------------------------------------------------------------

\end{document}